%% file: bslifetime-prl.tex
\documentclass[aps,prl,showpacs,twocolumn,groupedaddress,floatfix,nofootinbib]{revtex4}  
\usepackage{graphicx}  
\usepackage{dcolumn}   
\usepackage{bm}        
\usepackage{amssymb}   

\begin{document}

\hspace{5.2in} \mbox{FERMILAB-Pub-04/225-E}

\title{Measurement of the $B^{0}_{s}$ lifetime in the exclusive decay channel
$B^0_{s}\rightarrow J/\psi\phi$ }

\input list_of_authors_r2.tex

\date{\today}

\begin{abstract}
Using the exclusive decay   $B^0_{s} \rightarrow$ $J/\psi (\mu^{+}\mu^{-}) \phi(K^{+}K^{-})$, we report the most 
precise single  measurement of the $B^0_{s}$ lifetime.
The data sample  corresponds to an integrated luminosity of approximately  220~pb$^{-1}$ collected with the  D\O\ detector
at the Fermilab Tevatron Collider in 2002--2004.
We reconstruct 337 signal candidates, from which we extract the 
$B^0_s$ lifetime,
$\tau(B^0_s) = 1.444 ^{+0.098}_{-0.090} \thinspace (\mbox{stat}) \pm 0.020 \thinspace (\mbox{sys}) \thinspace \thinspace \mbox{ps}.$ 
We also report a measurement for the lifetime of the $B^0$ meson using the exclusive decay
$B^0\rightarrow J/\psi (\mu^{+}\mu^{-}) K^{*0}(892)(K^{+}\pi^{-})$. We reconstruct 1370 signal candidates, obtaining
$\tau(B^0) = 1.473 ^{+0.052}_{-0.050} \thinspace (\mbox{stat}) 
\pm 0.023 \thinspace (\mbox{sys})$~ps, and the ratio
of lifetimes,
$\tau(B^0_s)/\tau(B^0) = 0.980 ^{+0.076}_{-0.071} \thinspace (\mbox{stat}) \pm 0.003 \thinspace (\mbox{sys}).$
\end{abstract}


\maketitle

Lifetime differences among hadrons containing $b$ quarks can be used to probe decay
mechanisms that go beyond the quark-spectator model~\cite{B_life}. In the charm 
sector, lifetime differences are quite
large~\cite{cite:PDG}; however, in the bottom sector, due to the
larger $b$-quark mass, these differences are expected to be
smaller. Phenomenological models predict differences of about 5\%
between the lifetimes of $B^+$ and $B^0$, but no more than 1\% between 
$B^0$ and $B^0_s$ lifetimes~\cite{B_life}. These
predictions are consistent with previous measurements of $B$-meson
lifetimes~\cite{cite:PDG}. It has also been postulated~\cite{CP} that
the lifetimes of the two $CP$ eigenstates (of the 
$B^0_s$-$\bar{B^0_s}$ system)  differ. This could be observed as a difference in lifetime
between $B^0_s$ semileptonic decays, which should have an equal
mixture of the two   \mbox{\it CP\/}  eigenstates, and the lifetime for $B^0_s\rightarrow J/\psi\phi$, 
which is expected to be dominated by the \mbox{\it CP\/}-even eigenstate~\cite{CP}.

In this Letter, we report a measurement of the lifetime of the $B^0_{s}$ meson
using the exclusive decay channel $B^0_{s}\rightarrow J/\psi\phi$, followed by $J/\psi \rightarrow \mu^{+}\mu^{-}$  and $\phi
\rightarrow K^{+}K^{-}$. The lifetime is extracted using a
simultaneous unbinned maximum likelihood fit to
masses and proper decay lengths.  We also measure the lifetime
of the $B^0$ meson in the exclusive decay\footnote{Unless explicitly stated, the appearance of a specific charge state
will also imply its charge conjugate throughout this Letter.  } $B^0\rightarrow
J/\psi$ $K^{*0}(892)$, followed by $J/\psi \rightarrow \mu^{+}\mu^{-}$ and
$K^{*0}(892) \rightarrow K^{+}\pi^{-} $, and extract the ratio of the lifetimes
of the $B^0_s$ and $B^0$ mesons.
The analysis is based on data collected with the D\O\ detector in Run II of the Fermilab Tevatron Collider
during the period September 2002--February 2004, which corresponds to approximately 220~pb$^{-1}$ of 
$p\bar{p}$\ collisions at $\sqrt{s} = 1.96$~TeV.


The D\O\ detector is  described in detail elsewhere~\cite{run2det}.
We describe here only the detector components most relevant to this
analysis. The central-tracking system  
consists of a silicon microstrip tracker (SMT) and a central 
fiber tracker (CFT), both located inside a 2~T superconducting 
solenoidal magnet~\cite{run2det}. The tracking system and solenoid is 
surrounded by a liquid argon calorimeter. 
The SMT has $\approx 800,000$ 
individual strips, with typical pitch of $50-80$ $\mu$m, and a design 
optimized for tracking and vertexing capability  for   $|\eta|<3$,
where $\eta = -\ln[\tan(\theta/2)]$  is the pseudorapidity  and   $\theta$ 
is the polar angle measured  relative to the proton beam direction. 
The system has a six-barrel longitudinal structure, each with 
a set of four layers arranged axially around the beam pipe, 
and interspersed with sixteen radial disks. The CFT has eight thin 
coaxial barrels, each supporting two doublets of overlapping 
scintillating fibers of 0.835~mm diameter, one doublet 
parallel to the beam axis, and the other alternating by 
$\pm 3^{\circ}$ relative to this axis. Light signals are transferred 
via clear light fibers to solid-state photon counters  
that have  a quantum efficiency of approximately  $80\%$.
The muon system resides beyond the calorimeter, 
and consists of a layer of tracking detectors and scintillation 
trigger counters before 1.8~T toroidal magnets, followed by two  
similar layers after the toroids. Muon identification for $|\eta|<1$ relies 
on 10~cm wide drift tubes, while 1~cm wide mini-drift 
tubes are used for $1<|\eta|<2$. Coverage for muons is partially 
compromised at the bottom of the detector where the calorimeter is 
supported mechanically from the ground.
Luminosity is measured using plastic scintillator arrays located 
in front of the end  calorimeter cryostat, 
covering $2.7 < |\eta| < 4.4$. 

The data collection consists of a three-level
trigger system,  designed to accommodate the high luminosity 
of Run II. The first level uses  information
from the tracking, calorimetry, and muon systems to reduce the rate for 
accepted events to $\approx$ 1.5~kHz. At the next trigger level, 
with more refined information, the rate is reduced further to 
$\approx$ 800~Hz. The third and final level of the trigger, 
with access to all of the event information, uses software algorithms 
and a computing farm and reduces the output rate to $\approx$ 50~Hz, 
which is recorded   for further analysis.  We did not require the presence of  any  specific trigger 
in the event selection.

\begin{figure}[htb]
\begin{tabular}{c}
\includegraphics[width=3.5in]{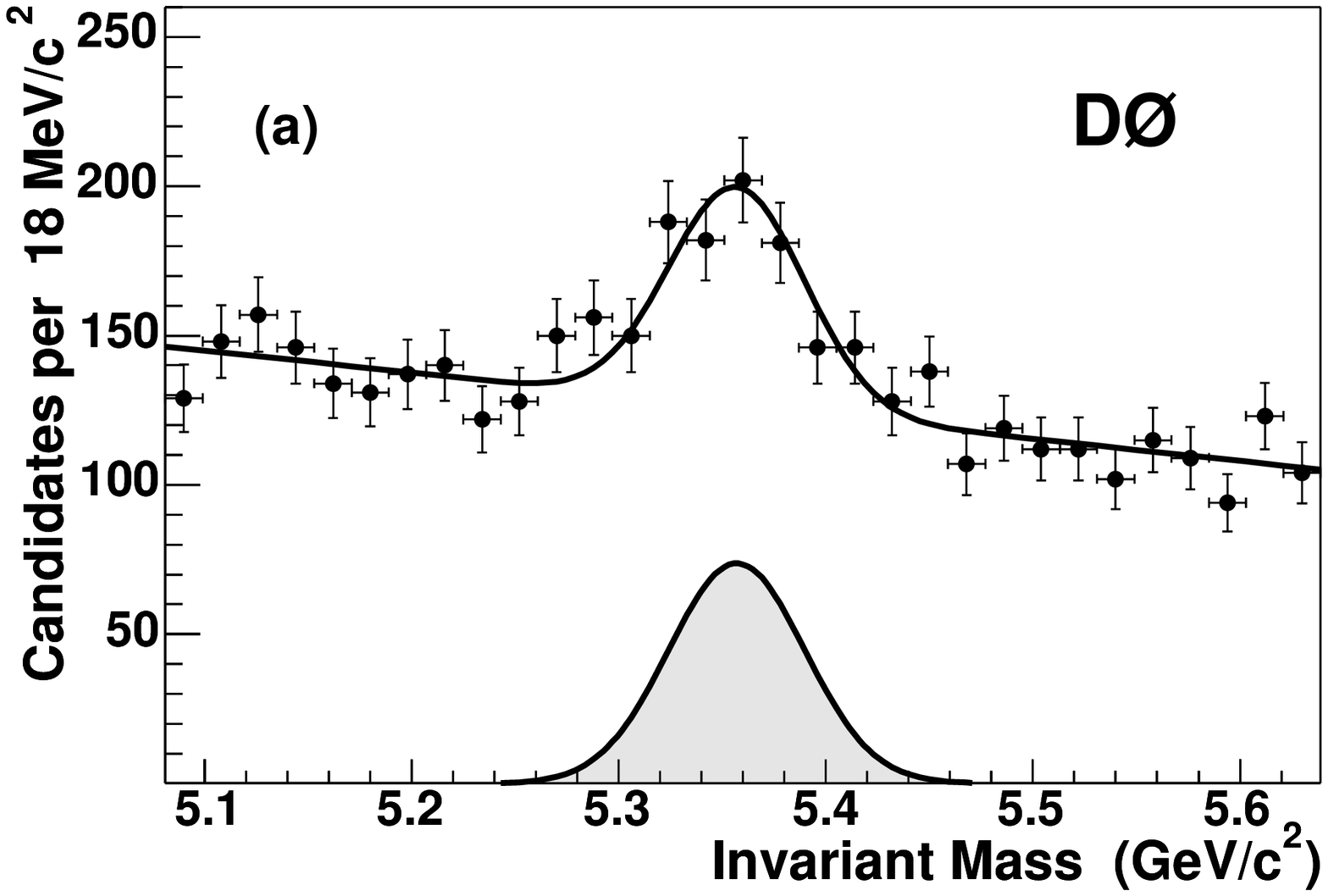} \\
\includegraphics[width=3.5in]{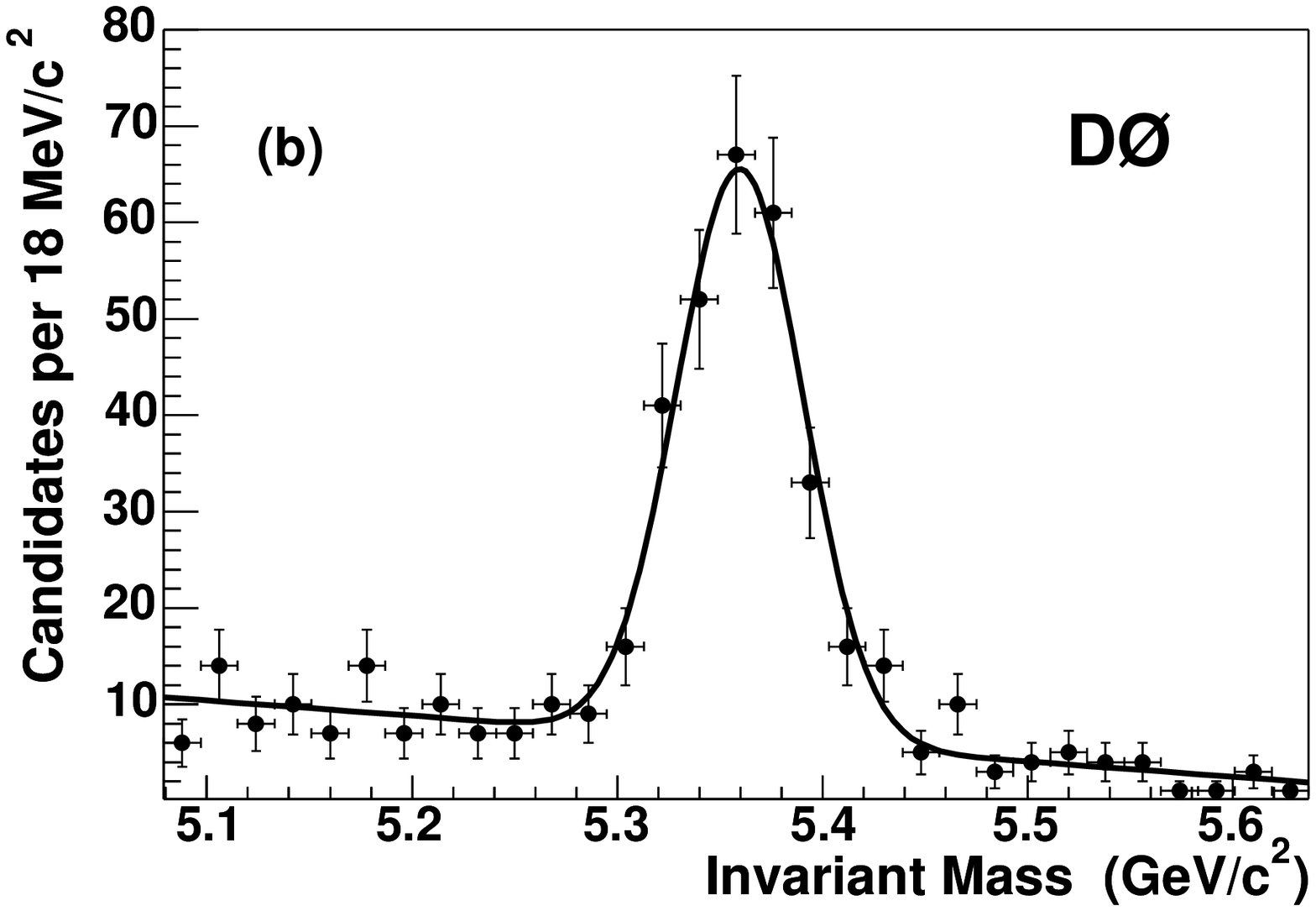}
\end{tabular}
\caption{\label{fig:uno} (a) Mass distribution
 for  $B_s^0$ candidate events. Points with error bars show the data, and the solid curve represents the 
result of the fit. The mass distribution for the signal is shown in gray;
(b) same distribution after requiring the
significance of the lifetime measurement to be  $c\tau/\sigma(c\tau) > 5$.}
\end{figure}

Reconstruction of $B^0_s \rightarrow J/\psi \phi$ candidates  requires 
a pair of oppositely charged muons that are identified by extrapolating charged 
tracks into the muon system and matching them with hits in the muon system.
All charged tracks used in this analysis are required to 
have at least one hit in the SMT.
We require that  muon candidates each have a minimum transverse momentum $p_T$ $>$ 1.5~GeV/$c$ 
and that they form a common vertex, according to the algorithm described in 
Ref.~\cite{PVref}, which is based on a  fit requiring a
$\chi^2$ probability greater than  1\%.
The dimuon system was required to have an invariant mass between 2.90 and 3.15 GeV/$c^{2}$ and 
transverse momentum above 4.5 GeV/$c$. The dimuons are then combined with another pair of
oppositely charged tracks, each with $p_T > $  0.8~GeV/$c$,
consistent with the decay $\phi \rightarrow K^+ K^-$.
The $\phi$ candidate was required to have an invariant mass between   1.008 and 1.032 GeV/$c^2$ 
and transverse momentum greater than 2~GeV/$c$. A four-track secondary vertex
is fitted to the products of the $J/\psi$ and $\phi$ decays, and required
to have a $\chi^2$ probability of at least 1\%.  The mass of the $J/\psi$ candidate is constrained 
in the fit to the world average
$J/\psi$ mass of 3.097 GeV/$c^2$~\cite{cite:PDG}, the constraint does not take into  
account the uncertainty in the  $J/\psi$ mass.  The
resulting $B^0_s$ candidate is required to have $p_T >$ 6.5 GeV/$c$.
We allow only one $B^0_s$ candidate per event, and when multiple candidates
exist, we choose the one with the best vertex probability.  
The resulting invariant mass distribution of the $J/\psi$-$\phi$
system is shown in Fig.~\ref{fig:uno}(a).

Each primary vertex is reconstructed using tracks and the mean beam-spot position. 
The latter is determined for every data run, where a
typical run lasts several hours.  
The initial primary vertex seed is constructed using all available tracks;
a track is removed when it causes a change of more than 9 units in the 
$\chi^2$ for a fit to a common vertex. 
The process is repeated until no more tracks can be removed~\cite{PVref}.

We take the four-track vertex as the position of the 
secondary vertex. To determine the distance traveled by each $B_s^0$ candidate, we calculate
the signed transverse decay length (in a plane transverse to the direction of the beam),
$L_{xy} = \vec{x} \cdot (\vec{p_T}/p_T)$, where $\vec{x}$ is the
 length vector pointing from the primary to the secondary vertex and $\vec{p_T}$
is the reconstructed transverse momentum vector of the $B_s^0$. 
The proper decay length of the $B_s^0$ candidate is
then defined as  $c\tau = L_{xy}(M_{B_s^0}/p_T)$,
where $M_{B_s^0}$ is taken as the  world average  mass of the 
$B_s^0$ meson  5.3696 GeV/$c^2$ ~\cite{cite:PDG}.

Figure~\ref{fig:uno}(b)  shows the reconstructed invariant mass
distribution of the $B^0_s$ candidates after a proper decay length
significance requirement of $c\tau/\sigma(c\tau) > 5$ is imposed, where
$\sigma(c\tau)$ is the uncertainty on $c\tau$. The strong suppression  of the background 
by this  cut  implies that  the  background is dominated by  zero lifetime vertices,  as expected.    

The proper decay length (without any restriction on significance) and the invariant mass distributions for
candidates passing the above criteria are fit simultaneously using an 
unbinned maximum likelihood method. 
The likelihood function ${\cal L}$ is given by:
\[{\cal L} = \prod^{N}_{i} [f_{s} {\cal F}_s^{i} + (1 - f_s) {\cal F}_b^{i}],\]
\noindent where ${\cal F}_s$ is the product of probability density functions for
mass and proper decay length for $B^0_s$,
${\cal F}_b$ is the equivalent for background,
$f_s$ is the fraction of signal, and $N$ is the
total number of candidate events in the sample.

The proper decay length for signal events is modeled by a normalized
exponential-decay function convoluted with a Gaussian function of width
equal to the uncertainty on the proper decay length, which is
typically $\approx$ 25~$\mu$m.  This uncertainty is
obtained from the full covariance (error) matrix of tracks at the
secondary vertex and the uncertainty in the position of the primary
vertex.  The uncertainty is multiplied by a scale factor that is
a parameter in the fit to allow for a possible misestimate of
the decay length uncertainty. The mass distribution of signal events
is modeled by a Gaussian function.

The proper decay length for the  background is parametrized as a
sum of a Gaussian function centered at zero and exponential
decay functions, with two short-lived components and a
long-lived term. The long-lived 
component accounts for heavy-flavor backgrounds, 
while the other terms account for
resolution and prompt contributions to background.  The mass distribution for the background
is modeled by a first-order polynomial.

\begin{figure}[htb]
\begin{tabular}{c}
\includegraphics[width=3.5in]{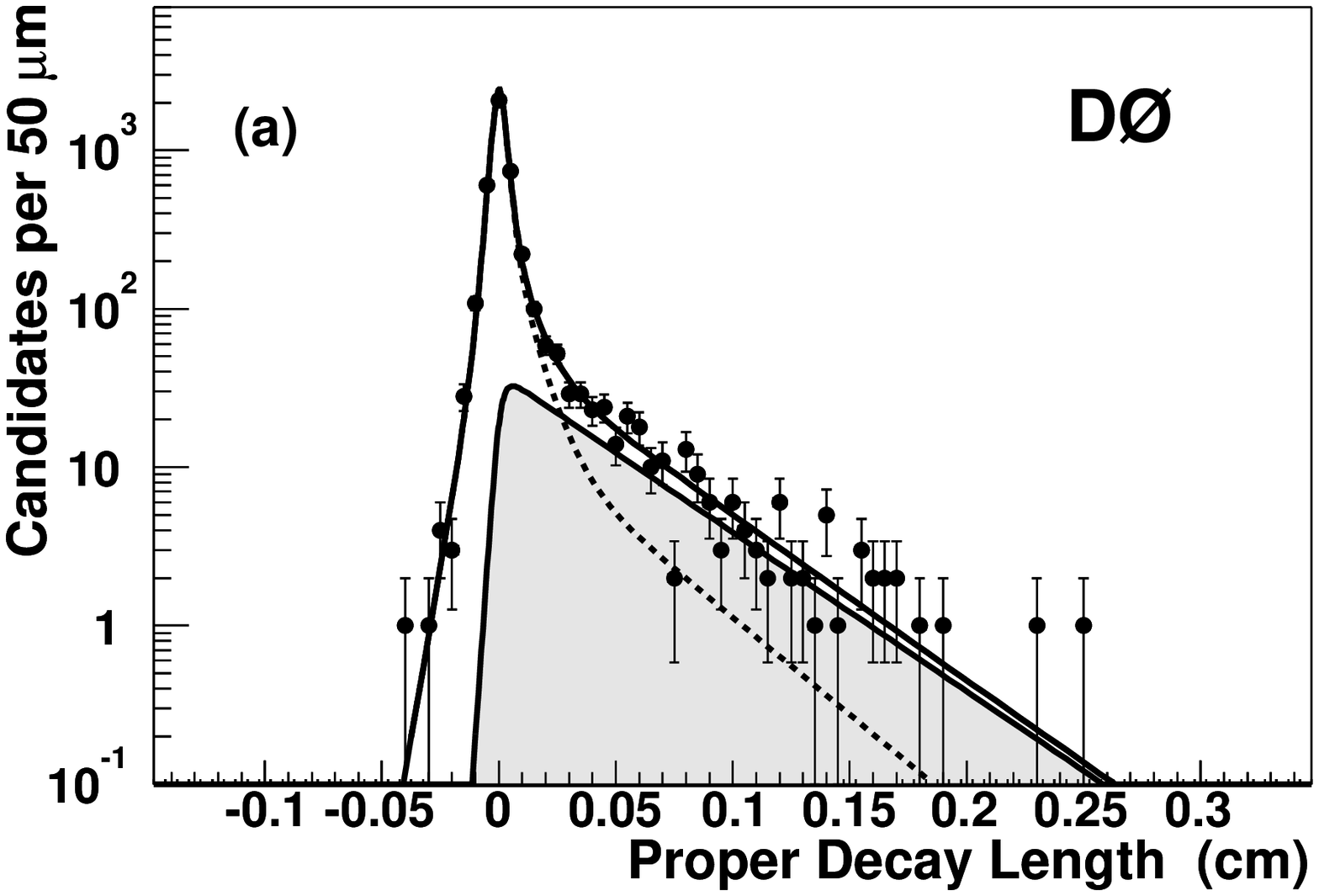}\\
\includegraphics[width=3.5in]{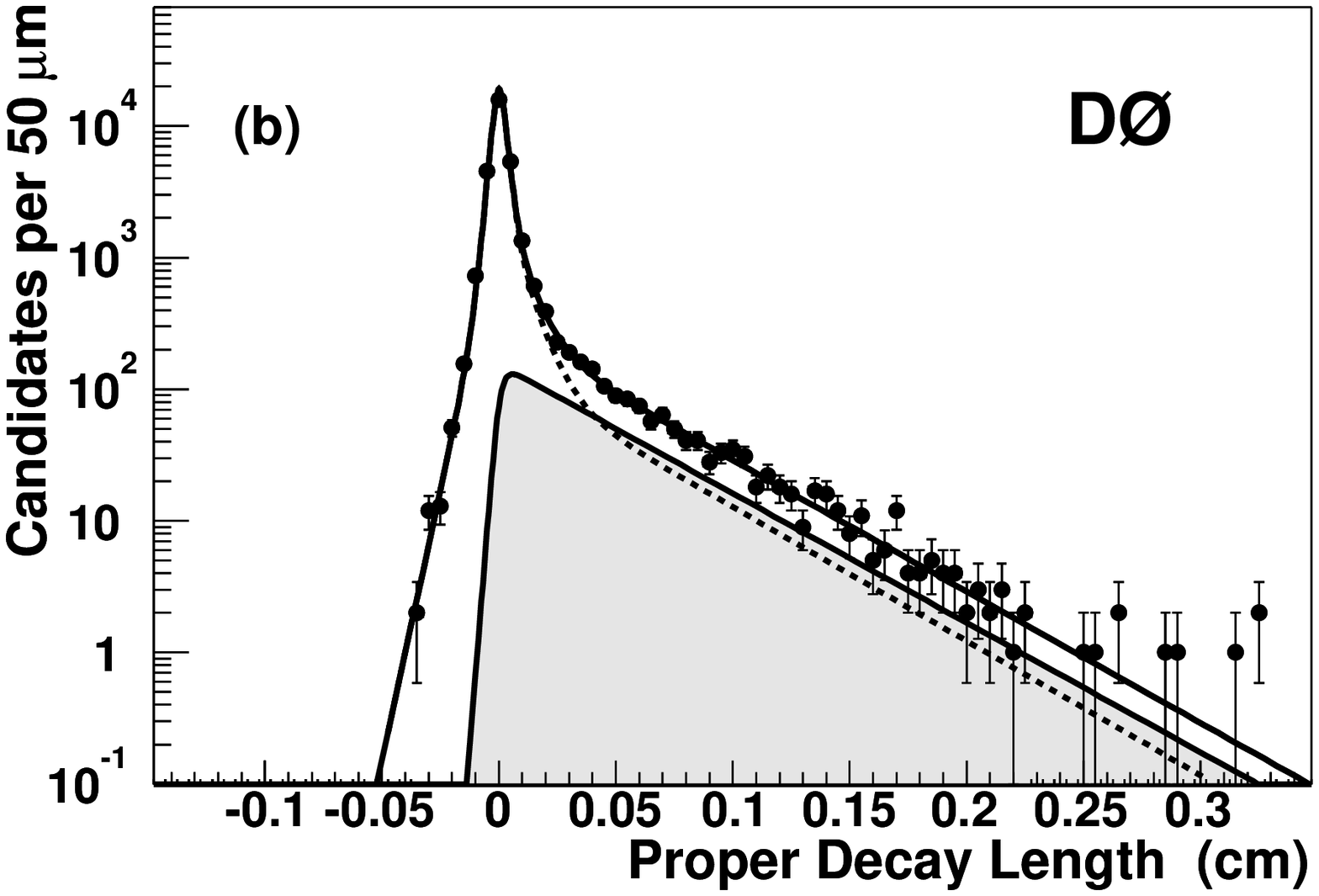}
\end{tabular}
\caption{\label{fig:tres} Proper decay length 
distributions for 
(a) $B_s^{0}$ 
and (b) $B^0$ 
candidates. The points with error bars show the data. 
The solid curve shows the total fit,
the dashed curve the background component, and the 
shaded region the signal.} 
\end{figure}

To determine the background we use a wide mass
range of 5.078--5.636 GeV/$c^2$ in the fit, corresponding to 4236 $B_s^0$ candidates. 
The number of background candidates in this range is sufficiently large to measure the
parameters of the background with high accuracy and therefore extract a
good measurement of the signal fraction and $c\tau(B_s^0)$.
The fit provides the  $c\tau$ and mass of the $B_s^0$, the shapes of
the proper decay length and mass distributions for the  background, and the
signal fraction.  Table~\ref{tab:bsfit} lists the fit
values of the parameters and their uncertainties.
The distribution of proper decay length and fits
to the $B^0_s$ candidates are shown in Fig.~\ref{fig:tres}(a).
  
\begin{table}[htb]
\caption{\label{tab:bsfit}Values of the extracted mass
$M_B$, resolution on the reconstructed mass $\sigma_M$, 
the measured  $c\tau$, the signal fractions $f_s$, and the scale factor $s$.}
\begin{ruledtabular}
\begin{tabular}{ccc}
Parameter   & $B_s^0\rightarrow J/\psi\phi$  & $B^0\rightarrow J/\psi K^{*0}(892)$ \\
            &  fit values                 & fit values                  \\ \hline
 $M_B$      & $5357.0 \pm 2.5$~MeV/c$^2$     & $5271.2 \pm 1.5$~MeV/c$^2$     \\ 
$\sigma_M$  & $32.9^{+2.5}_{-2.3}$~MeV/c$^2$ & $37.9^{+1.4}_{-1.3}$~MeV/c$^2$ \\  
$c\tau$ & $433^{+30}_{-27}$~$\mu$m & $442^{+16}_{-15}$~$\mu$m \\ 
 $f_s$      & $0.0796 \pm 0.0058$              & $0.0446 \pm 0.0018$              \\
$s$         & $1.142 \pm 0.028$              & $1.128 \pm 0.009$              \\
\end{tabular}
\end{ruledtabular}
\end{table}

With a very similar four-track topology in the final state, the
exclusive decay $B^0 \rightarrow J/\psi K^{*0}(892)$ followed by $J/\psi
\rightarrow \mu^+\mu^-$ and $K^{*0}(892) \rightarrow K^+ \pi^-$ is
reconstructed using the same selection criteria and algorithms as for
the $B^0_s$ channel described above.  The only differences are the
requirement that the $p_T$ of the pion be greater than 0.5~GeV/$c$, and
the selection of the $K^{*0}(892)$ candidates.  The combination of two
oppositely charged tracks, assuming the pion mass for one and the kaon mass for the other,
that gives an invariant mass closest to the mass of the $K^{*0}(892)$~\cite{cite:PDG} is 
selected for further study.  The invariant mass of these combinations is required 
to be between 0.850 and 0.930~GeV/$c^2$.  
Using the sample of $B^0$ candidates in the mass range
4.935--5.610 GeV/$c^2$, corresponding to 30692 candidates, we determine the  $c\tau$  and mass of the 
$B^0$ using exactly the same procedure as used for $B_s^0$ mesons.
Results are also given in Table~\ref{tab:bsfit}, and  the distribution of
proper decay length is shown in Fig.~\ref{fig:tres}(b).


Detailed Monte Carlo studies 
were performed on ensembles of events comparable to data samples,
with similar resolutions, pulls, fitting and selection criteria. No significant
biases resulting from our analysis procedures were observed.
To test the stability of the fit results for  
$B_s^0$ and $B^0$
mesons, we  split each data sample into two roughly equal parts in order to study
different kinematic and geometric parameters,  compared the fit
results, and found  consistency within their uncertainties.
We varied the selection criteria and mass ranges, and did not observe any significant shifts. 
Using Monte Carlo samples with different  
input proper decay lengths in the range 340 to 560 $\mu$m,  we checked the
response of our fits to this variation, and found it to be linear in this range.
We studied the contamination of our sample from cross-feed between
$B_s^0$ and $B^0$ using Monte Carlo events. The estimated contamination is
4.4\% for $B_s^0$ and 1.1\% for $B^0$, with invariant mass 
spread almost uniformly across the entire mass range.  Therefore, their
contributions  are included  in the long-lived heavy-flavor component of the background. 
To study possible biases from our fitting procedure,
we used toy Monte Carlo ensembles with the same  statistics as our data and with
distributions  matching those in data. These  
samples were fit, and the resulting means and widths of the distributions of 
extracted parameters are consistent with the fits to data.

Other sources of systematic uncertainty have been considered, and
the contributions are listed in Table~\ref{tab:systsum}.
For the $B^0_s$ lifetime, there are  
major contributions from determination of
the background, the model for resolution, and the reconstruction 
of the secondary vertex.
To determine the systematic error due to the uncertainty in the  background,
we considered different models for the mass 
and decay-length distributions.
In particular, to account for any model dependence on the invariant mass of misreconstructed
heavy-flavor hadrons, we fit the probability distributions separately 
in the
lower-mass and higher-mass side-band regions, and found the long-lived component to have different exponents.
Combining the two lifetime values for 
the long-lived components, we modified the functional form 
of the long-lived component for the global background 
in our fit. 
The two long-lived components were combined using a  weighting parameter $w=0.98^{+0.02}_{-0.36}$. 
This weighting parameter was varied by its uncertainty. The largest difference in the  $c\tau(B_s^0)$  observed
in these variations of background modeling was found to be 4~$\mu$m,
and is taken as the systematic uncertainty  due to this source.  The
effect of uncertainty in the proper decay length resolution  was studied by using
an alternative resolution function consisting of two Gaussian
functions (with the same mean but different width), resulting in a
difference in the fitted  $c\tau(B_s^0)$ of 3~$\mu$m. Uncertainty or biases in
the determination of the secondary vertex were estimated using secondary vertices
constructed with the  $J/\psi$ tracks only,
resulting in a  $c\tau(B_s^0)$  shift of 3~$\mu$m.
The contribution from the uncertainty on the detector alignment is estimated by
reconstructing  the $B^0_s$ candidate events with the position of the SMT sensors 
shifted radially outwards  by the alignment error  in the radial position of the sensors. 
The resulting difference in fitted  proper decay length of
2~$\mu$m is taken as the  systematic uncertainty  due to possible misalignment. 
The total systematic uncertainty from all these sources added in
quadrature is 6~$\mu$m.
The  systematic uncertainties 
in the measurement of the $c\tau(B^0)$ are determined in the same way
as for the $B_s^0$, and each contribution is listed in
Table~\ref{tab:systsum}.

\begin{table}[htb]
\caption{\label{tab:systsum}Summary of systematic uncertainties.}
\begin{ruledtabular}
\begin{tabular}{cccc}
                    & $c\tau(B_s^0) $ & $c\tau(B^0)$   & $\tau(B_{s}^0) /\tau(B^0) $ \\
                     & ($\mu$m)                      & ($\mu$m)                         &   \\
\hline
Alignment            &       2                       &           2                      &    0.000   \\
$J/\psi$\ vertex     &       3                       &           4                      &    0.002   \\
Model for resolution &       3                       &           3                      &    0.000   \\
Background           &       4                       &           5                      &    0.002   \\ \hline

{\bf Total}          &   {\bf 6}                     &         {\bf 7}                  & {\bf 0.003}   \\
\end{tabular}
\end{ruledtabular}
\end{table}


We determine the lifetimes of the $B_s^0$ and $B^0$ mesons,

\[\tau(B_{s}^0) = 1.444 ^{+0.098}_{-0.090}\ (\mbox{stat}) ~  
\pm 0.020\ (\mbox{sys})\ \mbox{ps,}\]

\[\tau(B^0) = 1.473 ^{+0.052}_{-0.050}\ (\mbox{stat}) 
\pm 0.023\ (\mbox{sys})\ \mbox{ps.}\]

Both results are consistent with the current world averages 
of $\tau(B_{s}^0) = 1.461 \pm 0.057$~ps and 
$\tau(B^0) = 1.536 \pm 0.014$~ps~\cite{cite:PDG}.
We note that measurements
using $B_{s}^0$ semileptonic events, where there is an equal
mixture of \mbox{\it CP\/}-even and \mbox{\it CP\/}-odd states, dominate the current world
average, while $B^0_{s} \rightarrow$ $J/\psi \phi $  has a different 
composition of \mbox{\it CP\/}-even and \mbox{\it CP\/}-odd states as discussed earlier \cite{citeCP}.

Using our results we determine the ratio of
$B_s^0/B^0$ lifetimes to be 

\[\frac{\tau(B_{s}^0) }{\tau(B^0) } = 0.980\ ^{+0.076}_{-0.071}\ (\mbox{stat})
\ \pm 0.003\ (\mbox{sys}),\]
where statistical uncertainties  were propagated 
in quadrature, and the systematic uncertainty was evaluated by adding each 
contribution to the corresponding central value, and evaluating a new 
ratio, with the difference from the  nominal value
taken as the systematic uncertainty of that source, as shown
in Table~\ref{tab:systsum}. The sum in quadrature of all contributions is 
reported as the overall systematic uncertainty on the ratio of lifetimes 
including  correlations between  the two lifetime measurements. 

In conclusion, we have measured the $B_s^0$ and $B^0$ lifetimes in exclusive
decay modes in $p\bar{p}$ collisions. The measurements are consistent 
with previous  results~\cite{cite:PDG}.
The value of the $B_s^0$ lifetime obtained in this analysis is 
the most precise measurement from any single experiment. 
The ratio of the lifetimes is also in 
good agreement with QCD models based on a heavy quark expansion, 
which predict 
a difference between $B_s^0$ and $B^0$ lifetimes 
of the order of 1\%~\cite{B_life}.

\input acknowledgement_paragraph_r2.tex


\end{document}

%% file: list_of_authors_r2.tex
%
\author{                                                                      
V.M.~Abazov,$^{33}$                                                           
B.~Abbott,$^{70}$                                                             
M.~Abolins,$^{61}$                                                            
B.S.~Acharya,$^{27}$                                                          
D.L.~Adams,$^{68}$                                                            
M.~Adams,$^{48}$                                                              
T.~Adams,$^{46}$                                                              
M.~Agelou,$^{17}$                                                             
J.-L.~Agram,$^{18}$                                                           
S.N.~Ahmed,$^{32}$
% ...
\\                                                                            
\vskip 0.30cm                                                                 
\centerline{(D\O\ Collaboration)}                                             
\vskip 0.30cm                                                                 
}                                                                             
\address{                                                                     
\centerline{$^{1}$Universidad de Buenos Aires, Buenos Aires, Argentina}       
\centerline{$^{2}$LAFEX, Centro Brasileiro de Pesquisas F{\'\i}sicas,         
                  Rio de Janeiro, Brazil}                                     
\centerline{$^{3}$Universidade do Estado do Rio de Janeiro,                   
                  Rio de Janeiro, Brazil}                                     
\centerline{$^{4}$Instituto de F\'{\i}sica Te\'orica, Universidade            
                  Estadual Paulista, S\~ao Paulo, Brazil}                     
\centerline{$^{5}$University of Alberta, Edmonton, Canada and Simon
Fraser University, Burnaby, Canada}
\centerline{$^{6}$Institute of High Energy Physics, Beijing,                  
                  People's Republic of China}                                 
\centerline{$^{7}$Universidad de los Andes, Bogot\'{a}, Colombia}             
\centerline{$^{8}$Charles University, Center for Particle Physics,            
                  Prague, Czech Republic}                                     
\centerline{$^{9}$Czech Technical University, Prague, Czech Republic}         
\centerline{$^{10}$Institute of Physics, Academy of Sciences, Center          
                  for Particle Physics, Prague, Czech Republic}               
\centerline{$^{11}$Universidad San Francisco de Quito, Quito, Ecuador}        
\centerline{$^{12}$Laboratoire de Physique Corpusculaire, IN2P3-CNRS,         
                 Universit\'e Blaise Pascal, Clermont-Ferrand, France}        
\centerline{$^{13}$Laboratoire de Physique Subatomique et de Cosmologie,      
                  IN2P3-CNRS, Universite de Grenoble 1, Grenoble, France}     
\centerline{$^{14}$CPPM, IN2P3-CNRS, Universit\'e de la M\'editerran\'ee,     
                  Marseille, France}                                          
\centerline{$^{15}$Laboratoire de l'Acc\'el\'erateur Lin\'eaire,              
                  IN2P3-CNRS, Orsay, France}                                  
\centerline{$^{16}$LPNHE, Universit\'es Paris VI and VII, IN2P3-CNRS,         
                  Paris, France}                                              
\centerline{$^{17}$DAPNIA/Service de Physique des Particules, CEA, Saclay,    
                  France}                                                     
\centerline{$^{18}$IReS, IN2P3-CNRS, Universit\'e Louis Pasteur,
Strasbourg, France and Universit\'e de Haute Alsace, Mulhouse, France}
\centerline{$^{19}$Institut de Physique Nucl\'eaire de Lyon, IN2P3-CNRS,      
                   Universit\'e Claude Bernard, Villeurbanne, France}         
\centerline{$^{20}$RWTH Aachen, III. Physikalisches Institut A,               
                   Aachen, Germany}                                           
\centerline{$^{21}$Universit{\"a}t Bonn, Physikalisches Institut,             
                  Bonn, Germany}                                              
\centerline{$^{22}$Universit{\"a}t Freiburg, Physikalisches Institut,         
                  Freiburg, Germany}                                          
\centerline{$^{23}$Universit{\"a}t Mainz, Institut f{\"u}r Physik,            
                  Mainz, Germany}                                             
\centerline{$^{24}$Ludwig-Maximilians-Universit{\"a}t M{\"u}nchen,            
                   M{\"u}nchen, Germany}                                      
\centerline{$^{25}$Fachbereich Physik, University of Wuppertal,               
                   Wuppertal, Germany}                                        
\centerline{$^{26}$Panjab University, Chandigarh, India}                      
\centerline{$^{27}$Tata Institute of Fundamental Research, Mumbai, India}     
\centerline{$^{28}$University College Dublin, Dublin, Ireland}                
\centerline{$^{29}$Korea Detector Laboratory, Korea University,               
                   Seoul, Korea}                                              
\centerline{$^{30}$CINVESTAV, Mexico City, Mexico}                            
\centerline{$^{31}$FOM-Institute NIKHEF and University of                     
                  Amsterdam/NIKHEF, Amsterdam, The Netherlands}               
\centerline{$^{32}$University of Nijmegen/NIKHEF, Nijmegen, The               
                  Netherlands}                                                
\centerline{$^{33}$Joint Institute for Nuclear Research, Dubna, Russia}       
\centerline{$^{34}$Institute for Theoretical and Experimental Physics,        
                  Moscow, Russia}                                             
\centerline{$^{35}$Moscow State University, Moscow, Russia}                   
\centerline{$^{36}$Institute for High Energy Physics, Protvino, Russia}       
\centerline{$^{37}$Petersburg Nuclear Physics Institute,                      
                   St. Petersburg, Russia}                                    
\centerline{$^{38}$Lund University, Lund, Sweden, Royal Institute of
Technology and Stockholm University, Stockholm, Sweden and }
\centerline{Uppsala University, Uppsala, Sweden}
\centerline{$^{39}$Lancaster University, Lancaster, United Kingdom}           
\centerline{$^{40}$Imperial College, London, United Kingdom}                  
\centerline{$^{41}$University of Manchester, Manchester, United Kingdom}      
\centerline{$^{42}$University of Arizona, Tucson, Arizona 85721}              
\centerline{$^{43}$Lawrence Berkeley National Laboratory and University of    
                  California, Berkeley, California 94720}                     
\centerline{$^{44}$California State University, Fresno, California 93740}     
\centerline{$^{45}$University of California, Riverside, California 92521}     
\centerline{$^{46}$Florida State University, Tallahassee, Florida 32306}      
\centerline{$^{47}$Fermi National Accelerator Laboratory, Batavia,            
                   Illinois 60510}                                            
\centerline{$^{48}$University of Illinois at Chicago, Chicago,                
                   Illinois 60607}                                            
\centerline{$^{49}$Northern Illinois University, DeKalb, Illinois 60115}      
\centerline{$^{50}$Northwestern University, Evanston, Illinois 60208}         
\centerline{$^{51}$Indiana University, Bloomington, Indiana 47405}            
\centerline{$^{52}$University of Notre Dame, Notre Dame, Indiana 46556}       
\centerline{$^{53}$Iowa State University, Ames, Iowa 50011}                   
\centerline{$^{54}$University of Kansas, Lawrence, Kansas 66045}              
\centerline{$^{55}$Kansas State University, Manhattan, Kansas 66506}          
\centerline{$^{56}$Louisiana Tech University, Ruston, Louisiana 71272}        
\centerline{$^{57}$University of Maryland, College Park, Maryland 20742}      
\centerline{$^{58}$Boston University, Boston, Massachusetts 02215}            
\centerline{$^{59}$Northeastern University, Boston, Massachusetts 02115}      
\centerline{$^{60}$University of Michigan, Ann Arbor, Michigan 48109}         
\centerline{$^{61}$Michigan State University, East Lansing, Michigan 48824}   
\centerline{$^{62}$University of Mississippi, University, Mississippi 38677}  
\centerline{$^{63}$University of Nebraska, Lincoln, Nebraska 68588}           
\centerline{$^{64}$Princeton University, Princeton, New Jersey 08544}         
\centerline{$^{65}$Columbia University, New York, New York 10027}             
\centerline{$^{66}$University of Rochester, Rochester, New York 14627}        
\centerline{$^{67}$State University of New York, Stony Brook,                 
                   New York 11794}                                            
\centerline{$^{68}$Brookhaven National Laboratory, Upton, New York 11973}     
\centerline{$^{69}$Langston University, Langston, Oklahoma 73050}             
\centerline{$^{70}$University of Oklahoma, Norman, Oklahoma 73019}            
\centerline{$^{71}$Brown University, Providence, Rhode Island 02912}          
\centerline{$^{72}$University of Texas, Arlington, Texas 76019}               
\centerline{$^{73}$Rice University, Houston, Texas 77005}                     
\centerline{$^{74}$University of Virginia, Charlottesville, Virginia 22901}   
\centerline{$^{75}$University of Washington, Seattle, Washington 98195}       
}                                                                             

%% file: acknowledgement_paragraph_r2.tex
%
We thank the staffs at Fermilab and collaborating institutions, 
and acknowledge support from the 
Department of Energy and National Science Foundation (USA),  
Commissariat  \` a l'Energie Atomique and 
CNRS/Institut National de Physique Nucl\'eaire et 
de Physique des Particules (France), 
Ministry of Education and Science, Agency for Atomic 
   Energy and RF President Grants Program (Russia),
CAPES, CNPq, FAPERJ, FAPESP and FUNDUNESP (Brazil),
Departments of Atomic Energy and Science and Technology (India),
Colciencias (Colombia),
CONACyT (Mexico),
KRF (Korea),
CONICET and UBACyT (Argentina),
The Foundation for Fundamental Research on Matter (The Netherlands),
PPARC (United Kingdom),
Ministry of Education (Czech Republic),
Natural Sciences and Engineering Research Council and 
WestGrid Project (Canada),
BMBF and DFG (Germany),
A.P.~Sloan Foundation,
Civilian Research and Development Foundation,
Research Corporation,
Texas Advanced Research Program,
and the Alexander von Humboldt Foundation.
%

%% file: bslifetime-prl.bbl
\begin{thebibliography}{99}
%
\bibitem[*]{lehner}
Visitor from University of Zurich, Zurich, Switzerland.
\bibitem[\dag]{przybycien}
Visitor from Institute of Nuclear Physics, Krakow, Poland.
%
\vskip 0.25cm


  \bibitem{B_life} M. B. Voloshin and M. A. Shifman, Sov. Phys. JETP {\bf 64},
	698 (1986); I. Bigi and N. G. Uraltsev, Phys. Lett.  B {\bf 280},
	271 (1992); I. Bigi, Nuovo Cimento A {\bf 109}, 713 (1996);
        F.~Gabbiani, A.~I.~Onishchenko and A.~A.~Petrov, arXiv:hep-ph/0407004.

  \bibitem{cite:PDG}  S. Eidelman {\sl et al.} 
	(Particle Data Group), Phys. Lett. B {\bf 592}, 1 (2004).

  \bibitem{CP}I. Bigi {\sl et al.} in ``B Decays'', 2nd edition, edited by 
	S. Stone, (World Scientific, Singapore 1994) p. 13 ; I. Dunietz,
	Phys. Rev. D {\bf 52}, 3048 (1995);  M. Beneke, G. Buchalla, I. Dunietz, Phys.
	Rev. D {\bf 54}, 4419 (1996).


  \bibitem{run2det} T. LeCompte and H. T. Diehl, ``The CDF and D\O\ Upgrades for Run II'',  Ann. Rev. Nucl. Part. Sci.  {\bf 50}, 71 (2000); V. Abazov, {\sl et al.}, in preparation for submission  to Nucl. Instrum. Methods Phys. Res. A.

  \bibitem{PVref} J. Abdallah {\sl et al.} (DELPHI Collaboration), 
	Eur. Phys. J. C {\bf 32}, 185 (2004).

  \bibitem{citeCP} The exact fraction of \mbox{\it CP\/}-even and \mbox{\it CP\/}-odd decays is unknown. In this analysis,
                                   from Monte Carlo, the relative efficiency for \mbox{\it CP\/}-even and \mbox{\it CP\/}-odd states is  0.99 
                                     $\pm $ 0.01.

\end{thebibliography}
